\documentclass[10pt,conference]{IEEEtran}
\IEEEoverridecommandlockouts

\usepackage{cite}

\usepackage{textcomp}

\usepackage{multirow}
\usepackage[table,xcdraw]{xcolor}
\usepackage{tabularx} 
\usepackage{tcolorbox} 
\usepackage{amsfonts}
\usepackage{mathrsfs}
\usepackage{algorithm}          
\usepackage{algpseudocode}     
\usepackage{amsmath}            
\usepackage{graphicx}

\usepackage{subcaption} 
\usepackage{enumitem}
\usepackage{xurl}
\usepackage{amssymb}  
\usepackage[colorlinks=false, allcolors=blue, pdfborder={0 0 0}]{hyperref}

\def\BibTeX{{\rm B\kern-.05em{\sc i\kern-.025em b}\kern-.08em
    T\kern-.1667em\lower.7ex\hbox{E}\kern-.125emX}}
\begin{document}

\title{Weaponizing Ground Truth: Data Poisoning Attacks by Exploiting Boundary Misalignment Between Antivirus Software and Learning-Based Detectors}
\author{Jieshuai~Yang,
    Zhi~Wang,
    Yan~Jia,
    Zhenhua~Wu,
    Jianfei~Tang,
    Chenbin~Su,
    Jingwei~Ye,
    Jianwen~Tian,
    and Wanpeng~Li%
    \thanks{(Corresponding author: Zhi~Wang.)}%
    \thanks{Jieshuai~Yang, Zhi~Wang, Yan~Jia, Zhenhua~Wu,
    Jianfei~Tang, Chenbin~Su, and Jingwei~Ye are with the
    College of Cryptology and Cyber Science, Nankai University,
    Tianjin 300350, China
    (e-mail: yjs@mail.nankai.edu.cn;
    zwang@nankai.edu.cn;
    jiay@nankai.edu.cn;
    2120250720@mail.nankai.edu.cn;
    kid519388073@mail.nankai.edu.cn;
    champion.su@mail.nankai.edu.cn;
    jwye@mail.nankai.edu.cn).}%
    \thanks{Jianwen~Tian is with the School of Computing and
    Information Systems, Singapore Management University,
    80 Stamford Road, Singapore 178902
    (e-mail: jwtian@smu.edu.sg).}%
    \thanks{Wanpeng~Li is with the School of Computer Science
    and Informatics, University of Liverpool, Liverpool, UK
    (e-mail: wanpeng.li@liverpool.ac.uk).}%
}

\maketitle

\begin{abstract}
Machine-learning (ML)-based malware detectors are commonly trained using labels obtained from antivirus (AV) engines and aggregation services (e.g., VirusTotal). This practice assumes AV-generated labels provide reliable supervision. However, small byte-level modifications can substantially alter AV verdicts while leaving the representations perceived by downstream ML detectors largely unchanged, producing label-feature inconsistencies that can contaminate training datasets and create poisoning opportunities for ML-based malware detection.
We present Bi-Iocane, a black-box poisoning framework that exploits the reliance of malware-labeling pipelines on AV-generated labels. Bi-Iocane identifies AV-sensitive bytes and modifies them to induce label changes. It rewrites such bytes in malware to obtain benign labels (evasion-oriented poisoning) and injects malware-associated byte patterns into benign software to obtain malicious labels (defamation-oriented poisoning). These poisoned samples and their lightly modified variants corrupt training data and cause selected targets to be misclassified.
We evaluate Bi-Iocane with 13 AV engines simulating AV aggregation services and eight ML detectors. For 30 malware and 30 benign clean targets, Bi-Iocane combines AV-specific manipulations to generate malware-to-benign and benign-to-malware poisoned samples whose all tested AV-based labels are flipped. After these poisoned samples and variants are used for downstream training, the resulting ML models misclassify 92.08\% of the original clean targets on average with only a 0.06\% poisoning budget per target. Meanwhile, the poisoned models largely preserve clean-set performance, and six evaluated poisoning defenses show only limited mitigation. VirusTotal evaluation further confirms practical defamation risk and reveals potential evasion risk in real-world AV-to-ML labeling supply chains.

\end{abstract}


\section{Introduction}
As malware continues to grow and evolve~\cite{avtest2025malware}, traditional signature-based detection faces increasing difficulty in handling unseen and rapidly changing threats. This limitation has motivated ML-based malware detectors, which learn discriminative and generalizable patterns from large-scale historical labeled datasets and are better suited to detecting previously unseen or emerging malware families~\cite{bensaoud2024survey}.

However, ML models critically depend on the quality of their training labels. In practice, dataset builders and ML detector developers often obtain samples and labels from security vendors or multi-engine aggregation services, such as VirusTotal~\cite{virustotal} and VirScan~\cite{virscan}, and treat AV verdicts as ground truth for downstream training~\cite{anderson2018ember,harang2020sorel,yang2021bodmas,virustotal,virscan}. These users become primary victims when attacker-submitted samples enter the labeling supply chain with manipulated AV-assigned labels. Although this workflow enables scalable dataset construction, it inherits a fundamental mismatch between upstream AV engines and downstream ML detectors. AV engines still rely substantially on discrete signatures and patterns~\cite{wressnegger2017automatically,clamavsignatures,yara}, whereas ML detectors learn continuous and aggregated representations. Consequently, small byte-level perturbations can flip AV labels while largely preserving ML-perceived representations, making AV-to-ML labeling pipelines vulnerable to boundary-induced label manipulation.

Existing poisoning attacks against ML-based malware detectors mainly target downstream models rather than upstream label sources. Most are designed around the target model's input representation or feature extraction pipeline, modifying artifacts that affect detector-specific features such as API calls, byte sequences, section contents, or handcrafted statistical attributes~\cite{sasaki2019embedding,narisada2020stronger,severi2021explanation,yang2023jigsaw,zhan2025practical,li2021backdoor,zhang2023universal}. Even problem-space attacks still rely on such detector-specific feature effects.
Although effective in controlled settings, these attacks require the adversary to know, infer, or approximate the target representation, which is difficult for proprietary black-box detectors. The diversity of downstream representations, including handcrafted statistics, raw bytes, opcode sequences, graphs, and images, further limits their transferability. In contrast, manipulating AV-assigned labels in the upstream AV-to-ML pipeline offers a more general poisoning path for heterogeneous downstream detectors.

To the best of our knowledge, AndroVenom~\cite{lan2026trust} is the first work to shift poisoning from downstream detectors to the upstream AV-based labeling stage. It injects malicious files into benign Android APKs, causing AV engines or multi-engine services to assign malicious labels that later propagate into downstream training data. However, AndroVenom still relies on knowledge of the downstream feature space. Specifically, the injected files are placed in the APK \texttt{res} folder because widely used feature extractors, such as Drebin~\cite{arp2014drebin} and MaMaDroid~\cite{mariconti2016mamadroid}, do not process resource-file contents. The payload is therefore visible to AV engines but invisible to these downstream representations. If the target detector incorporates \texttt{res} contents, this assumption no longer holds. Thus, AndroVenom depends on a known downstream feature blind spot, is not fully feature-agnostic, and supports only unidirectional defamation, leaving reverse evasion unexplored.

To move beyond these limitations and expose a general supply-chain weakness, we propose \textbf{Bi-Iocane}, a downstream-detector-agnostic black-box poisoning framework that supports bidirectional attacks. Unlike AndroVenom, which relies on a known blind spot of downstream Android feature extractors, Bi-Iocane assumes no detector-specific invisible region or feature representation. Instead, it operates through the AV labeling interface and exploits decision-boundary misalignment between AV engines and ML-based detectors. By introducing subtle, localized perturbations to executable files, Bi-Iocane flips AV-assigned labels while largely preserving the representations perceived by downstream ML detectors.

Bi-Iocane constructs poisoned samples in two directions. For \textit{evasion}, it corrupts AV-sensitive boundary bytes in malware, causing AV engines to assign benign labels. For \textit{defamation}, it implants malware-mimicking boundary byte sequences into benign software to trigger malicious AV labels. The resulting samples and lightweight variants carry AV-assigned ``ground-truth'' labels and enter downstream training data, forcing detectors to associate nearly unchanged representations with flipped labels. At deployment time, the poisoned model misclassifies the original targets, either defaming benign software or enabling malware to evade detection. Since Bi-Iocane requires no knowledge of the downstream detector's architecture, feature extraction pipeline, or input representation, it provides a general poisoning pathway across heterogeneous ML-based malware detectors.

The results show that Bi-Iocane effectively poisons downstream ML-based malware detectors through AV-based labeling services. Across 13 AV engines, it achieves an average malware-to-benign label reversal rate of 90.96\% with only 0.09\textpertenthousand{} perturbation. For benign-to-malware reversal, Bi-Iocane identifies effective malware-mimicking byte sequences for 10 of 13 AV engines and achieves an average label reversal rate of 85.96\% with 0.13\% perturbation. We further use these engines to construct a controlled multi-AV labeling setting approximating VirusTotal and assume that eight downstream detectors across six feature families are trained using their labels. Under this 13-AV setting, Bi-Iocane achieves a 92.08\% average attack success rate with only a 0.06\% per-target poisoning ratio. The poisoned models preserve clean-set performance, with a maximum average F1 decrease of 0.15 percentage points, and retain an 82.14\% average attack success rate under the strongest of six evaluated defenses. To assess real-world impact, we submit 30 benign samples injected with malware-mimicking byte sequences and 30 malware samples rewritten at AV-sensitive byte positions to VirusTotal. The results show that malware-mimicking injection can push benign samples beyond common malicious-label thresholds, confirming practical defamation poisoning, while rewriting AV-sensitive bytes substantially lowers malicious AV verdicts, revealing a potential evasion-oriented poisoning risk.
In summary, this paper makes the following contributions:

\begin{itemize}

\item We identify an upstream label-supply-chain attack surface in ML-based malware detection. Attackers can exploit AV-to-ML boundary misalignment to inject samples with manipulated AV labels into downstream training data, causing targeted misclassification without knowing the target model.

\item We propose \textbf{Bi-Iocane}, a downstream-detector-agnostic black-box poisoning framework supporting bidirectional attacks. It rewrites AV-sensitive boundary points in malware for evasion and injects malware-mimicking boundary bytes into benign software for defamation.

\item We evaluate Bi-Iocane on 13 AV engines and eight ML-based detectors across six feature families. Under a controlled 13-AV setting, it achieves 92.08\% average attack success with a 0.06\% per-target poisoning ratio, while preserving clean-set performance and resisting existing defenses. VirusTotal validation confirms practical defamation risk and reveals potential evasion risk for multi-AV labeling pipelines.

\end{itemize}

The associated poisoning sample generation tool and evaluation implementation are available at 
\url{https://anonymous.4open.science/r/Bi-Iocane-25A7/}.

\section{Preliminaries and Threat Model}

\subsection{Preliminaries}

\label{Section:Preliminaries}
\textbf{Antivirus Software.}
Modern antivirus software may incorporate multiple detection mechanisms, but static signature- and pattern-based detection remains foundational due to its efficiency, low false-positive rate, and operational maturity~\cite{roy2025breaking}. 
These advantages make AV scanning suitable for large-scale malware dataset labeling, where massive files must be processed automatically and consistently. 
Pattern-based detection relies on signature databases built from manually analyzed or automatically generated malware patterns~\cite{naik2020evaluating,wressnegger2017automatically, clamavsignatures, yara}. 
When a file matches a known signature or byte pattern, the AV engine assigns a malicious verdict that can serve as a downstream training label.

\textbf{Machine Learning Detector.}
ML-based malware detectors usually learn statistical decision boundaries from feature representations rather than matching explicit byte-level signatures. 
Unlike AV engines that may rely on localized signature bytes, ML-based detectors typically make decisions based on aggregated feature representations extracted from broader byte regions, program structures, or statistical properties of the sample. 
We discuss representative ML-based malware detectors in Section~\ref{Section:Related_Work}.

\textbf{Decision Boundary Misalignment.}
As discussed above, AV engines and ML-based detectors define substantially different decision boundaries. 
The boundary of an AV engine is governed by a large set of discrete and deterministic pattern-matching rules accumulated over time. 
By contrast, the boundary of an ML-based detector is shaped by continuous and aggregated feature representations that capture software characteristics at different abstraction levels. 
This discrepancy leads to an important consequence. 
A software sample can be slightly modified to disrupt AV-sensitive patterns or inject malware-mimicking patterns, thereby flipping its AV-assigned label. 
For the ML-based detector, however, these localized modifications often have limited impact on its aggregated feature representation, leaving the modified sample close to the original in the ML feature space. 
This inherent boundary misalignment is the core vulnerability exploited in this work. 
By manipulating the AV decision boundary while largely preserving the ML-perceived representation, an attacker can weaponize AV-assigned ``ground-truth'' labels to silently poison the training data of downstream ML-based detectors.

\textbf{Label Sources for ML-based Malware Detectors.}
Existing ML-based malware detectors commonly inherit labels from security vendors or multi-AV aggregation services. The Microsoft Malware Classification Challenge is a representative single-vendor label source and has been used by models such as Feature-Fusion and MAlign~\cite{ronen2018microsoft,ahmadi2016novel,saha2024malign}. In contrast, EMBER derives its labels from VirusTotal detection results and has been used to train or evaluate LightGBM, MalConv, and MalConv2~\cite{anderson2018ember,raff2017malware,raff2021classifying}. Other detectors, such as MalGraph~\cite{ling2022malgraph}, also explicitly construct labels by aggregating VirusTotal reports~\cite{ling2022malgraph}.

For datasets with less transparent labeling pipelines, BODMAS states that most family labels are obtained from the verdicts of multiple AV vendors, while SOREL-20M provides labels derived from multiple security sources together with vendor-detection information~\cite{yang2021bodmas,harang2020sorel}. These examples show that aggregated AV labels, particularly those provided by platforms such as VirusTotal, are widely propagated into downstream ML training pipelines.
However, VirusTotal allows ordinary users to submit files for scanning, while privileged users and data-feed subscribers can access analyzed samples and reports~\cite{virustotalUploadFile,virustotalSearching,virustotalFileDownload}.
Consequently, attacker-submitted label-flipped samples may later be collected by dataset builders and enter downstream ML training sets with AV-aggregated labels.

\subsection{Threat Model}

\begin{figure}[t]\centering

	\centering
	\includegraphics[width=1\linewidth]{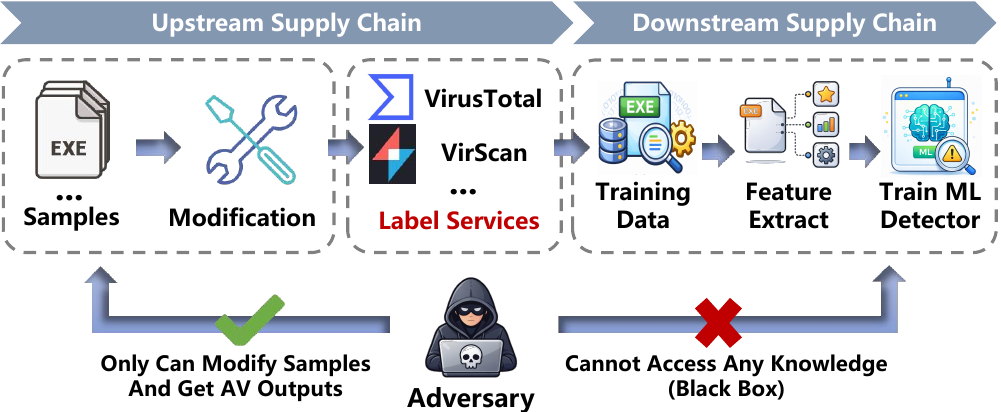}

\caption{Overview of the black-box threat model. The adversary only modifies samples submitted to AV-based labeling services and observes their outputs. Downstream users collect these samples and labels to train ML detectors, while the downstream pipeline remains unknown to the adversary.}

	\label{fig:Threat Model}
\end{figure}

\textbf{Assumptions.}
As shown in Figure~\ref{fig:Threat Model}, we consider an AV-to-ML data supply chain in which downstream ML-based malware detectors obtain training samples and labels from AV labeling services or datasets derived from them. The adversary can submit modified samples to the upstream labeling service and observe the resulting AV-assigned labels. However, the adversary has no knowledge of the downstream detector's architecture, weights, training data, data distribution, feature representation, or training procedure. The entire downstream pipeline is therefore treated as a black box.

\textbf{Victims and Delivery Path.}
The primary victims are ML detector developers or dataset builders who collect training samples and labels from a single AV engine or multi-AV aggregation services such as VirusTotal. The adversary submits label-flipped samples through normal public channels, which may later be collected with their AV-assigned labels and incorporated into downstream training data, without requiring direct access to the victim's training pipeline.

\textbf{Adversary Capabilities.}
The adversary can apply subtle perturbations to executable files to flip their AV verdicts while largely preserving their downstream ML representations. The resulting samples therefore carry AV-assigned ``ground-truth'' labels that conflict with the characteristics perceived by the downstream detector.

\textbf{Adversary Goals.}
The adversary aims to induce targeted misclassification of the original, unmodified samples after the victim trains the ML-based detector. This includes two directions: (1) \textit{evasion}, where target malware is misclassified as benign, harming the detector operator and protected users; and (2) \textit{defamation}, where target benign software is misclassified as malicious, harming its developer, vendor, or users. Meanwhile, the attack should preserve overall clean-set performance so that the poisoning remains difficult to detect during validation or deployment.

\begin{figure*}[t]\centering
	\centering
	\includegraphics[width=1\linewidth]{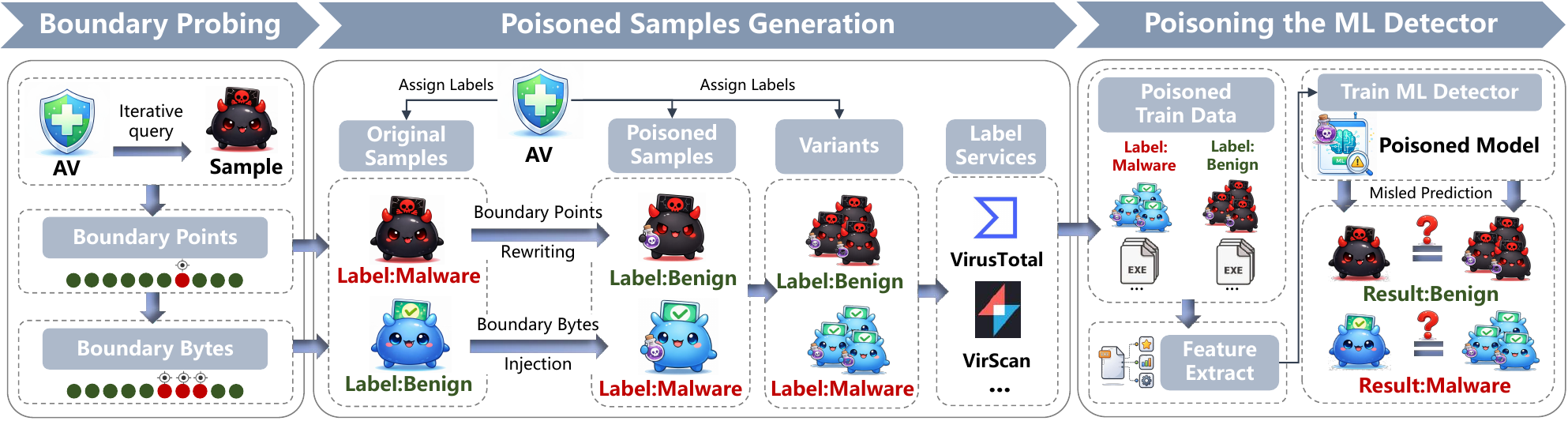}

\caption{Workflow of Bi-Iocane: (1) probe specific AV BPs and infer BBs; (2) generate label-flipped poisoned samples and their variants, and upload them to AV-based labeling services; (3) victims collect these samples and labels for training, poisoning downstream ML detectors to misclassify the original targets.}


	\label{fig:Workflow}
    
\end{figure*}

\section{Approach}


The overall workflow of Bi-Iocane is illustrated in Figure~\ref{fig:Workflow}. 
Bi-Iocane first probes malware samples to extract boundary points (BPs) and infer reusable boundary bytes (BBs). 
It then rewrites BPs in malware or injects BBs into benign software to generate label-flipped poisoned samples, together with multiple lightweight variants. 
These samples are uploaded to AV-based labeling services and receive manipulated AV-assigned labels. 
Victims relying on these services may subsequently collect the samples and labels into downstream training datasets. 
Training on the poisoned data causes ML-based detectors to misclassify the original targets, enabling malware evasion and benign-software defamation.

\subsection{Boundary Probing from Upstream AV Engines}

To precisely manipulate AV verdicts, Bi-Iocane first identifies the critical byte positions in a binary that determine whether it is classified as malicious or benign. 
We design a binary-search-based probing algorithm to extract BPs and BBs from a target AV engine. 
BPs refer to key byte positions whose modification can disrupt the corresponding AV detection pattern, while BBs denote reusable byte sequences that can trigger malicious AV verdicts when injected into benign samples. 
The detailed probing process is described below.

\subsubsection{Boundary Point Probing}
\label{Section:Boundary Points Probing}

Given a malware sample $m$ that is initially detected as malicious by the target AV, we take the entire byte range of the file as the search interval $[0, L-1]$, where $L$ is the file size in bytes. 
An empty set $\mathcal{BP}$ is initialized to store the discovered BPs.

Bi-Iocane locates BPs through an iterative binary-search procedure. Starting from interval $[low, high]=[0,L-1]$, it computes $mid=\lfloor(low+high)/2\rfloor$ and generates a modified sample $m'$ by overwriting bytes in $[mid,high]$ with \texttt{0x90}; bytes already equal to \texttt{0x90} are replaced with random alternatives to ensure effective perturbation. The sample is submitted to the AV, producing $r\in\{0,1\}$, where $0$ and $1$ denote benign and malicious, respectively. If $r=0$, the critical byte lies in the overwritten region and the interval becomes $[mid,high]$; otherwise, it becomes $[low,mid-1]$. The search repeats until one byte remains, which is recorded as a BP $b$ and permanently masked in the base sample. Bi-Iocane then repeats the same search to find additional, possibly non-contiguous BPs, until the masked sample is classified as benign or the maximum number of BPs is reached. Adjacent BPs are finally merged into contiguous boundary regions, and the resulting set $\mathcal{BP}$ is stored for subsequent steps.

\subsubsection{Boundary Byte Probing}

For each detected boundary point $b \in \mathcal{BP}$, we extract a contiguous byte sequence centered at $b$ as a reusable trigger pattern, termed boundary bytes (BBs). 
The objective is to find the shortest byte sequence that can trigger malicious AV verdicts with a sufficiently high success rate when injected into benign samples. 
As defined in Eq.~\ref{eq:bb_definition}, let $S_m(b,w)$ denote the byte sequence of length $2w+1$ centered at $b$ in malware sample $m$, where the indices are clipped to the file bounds. 
Let $\mathrm{Inject}(x,S)$ denote the operation of embedding a byte sequence $S$ into a benign sample $x$, and let $f_{\mathrm{AV}}(\cdot) \in \{0,1\}$ denote the AV verdict, where $1$ indicates malicious and $0$ indicates benign.

\begin{equation}
\label{eq:bb_definition}
S_m(b,w) = m[b-w : b+w+1],
\end{equation}

To evaluate the effectiveness of a candidate BB, we inject it into a benign validation set $\mathcal{X}_{\mathrm{v}}$ and compute its triggering rate. 
During injection, we prioritize section caves and non-code sections to avoid modifying executable code whenever possible, thereby minimizing disruption to the sample's functionality and code structure. 
As shown in Eq.~\ref{eq:bb_trigger_rate}, a candidate sequence is accepted if its triggering rate reaches a predefined threshold $\tau$.

\begingroup
\scriptsize
\begin{flalign}
&\mathrm{TR}(S)=
\frac{1}{|\mathcal{X}_v|}
\sum_{x\in\mathcal{X}_v}
\mathbf{1}\!\left[f_{\mathrm{AV}}(\mathrm{Inject}(x,S))=1\right],
\mathrm{TR}(S_m(b,w))\ge\tau . &&
\label{eq:bb_trigger_rate}
\end{flalign}
\endgroup

In our implementation, we set $|\mathcal{X}_{\mathrm{v}}|=10$ and use $\tau=0.5$ by default. 
To obtain a compact BB, we search for the minimum window size $w$ that satisfies Eq.~\ref{eq:bb_trigger_rate}. 
Similar to BP probing, this search is performed through binary search over the range $[0,W_{\max}]$. 
The resulting BB is accepted only if its length is below the predefined limit $L_{\max}$, which is set to 1024 bytes by default. 
All accepted BBs are added to $\mathcal{BB}$ and later used for benign-to-malware label reversal.

\begin{algorithm}[t]
\caption{Boundary Probing and Boundary Bytes Extraction}
\label{alg:simple_probing}
\begin{algorithmic}[1]
\Require Malware sample $m$, target AV $f_{\mathrm{AV}}$, benign validation set $\mathcal{X}_{\mathrm{v}}$, trigger threshold $\tau$, size limit $L_{\max}=1024$, maximum number of boundary points $N_{\max}$
\Ensure Boundary point set $\mathcal{BP}$, boundary byte set $\mathcal{BB}$

\State $\mathcal{BP} \gets \emptyset$, $\mathcal{BB} \gets \emptyset$
\State $m_{\mathrm{work}} \gets m$ \Comment{working copy with discovered BPs masked}

\While{$f_{\mathrm{AV}}(m_{\mathrm{work}})=1$ and $|\mathcal{BP}| < N_{\max}$}
    \State $b \gets \mathrm{BinarySearchBP}(m_{\mathrm{work}}, f_{\mathrm{AV}})$
    \If{$b$ is not found}
        \State \textbf{break}
    \EndIf
    \State $\mathcal{BP} \gets \mathcal{BP} \cup \{b\}$
    \State $m_{\mathrm{work}} \gets \mathrm{Mask}(m_{\mathrm{work}}, b)$
    \Comment{set byte $b$ to \texttt{0x90}}
\EndWhile

\For{each $b \in \mathcal{BP}$}
    \State Use binary search over $w \in [0,W_{\max}]$ to find the smallest $w$ such that
    \State \quad $\mathrm{TR}(m[b-w:b+w+1]) \geq \tau$ and $|m[b-w:b+w+1]| < L_{\max}$
    \If{such $w$ exists}
        \State $\mathcal{BB} \gets \mathcal{BB} \cup \{m[b-w:b+w+1]\}$
    \EndIf
\EndFor

\State \Return $\mathcal{BP}, \mathcal{BB}$
\end{algorithmic}
\end{algorithm}

\subsection{Poisoned Sample Generation}

Bi-Iocane generates poisoned samples in two directions.

\subsubsection{Poisoned Malware Sample Generation}
\label{Section:Poisoned Malware Sample Generation}

For malware originally detected as malicious, Bi-Iocane obtains its BP set $\mathcal{BP}$ from the target AV and overwrites each position $b \in \mathcal{BP}$ with the neutral byte \texttt{0x90}. This corrupts AV-sensitive boundary bytes, disrupts the corresponding detection patterns, and causes the modified sample to be classified as benign. The resulting file is recorded as a poisoned malware sample.

\subsubsection{Poisoned Benign Sample Generation}
\label{Section:Poisoned Benign Sample Generation}

To generate a poisoned benign sample, Bi-Iocane builds a pool of effective BBs, denoted as $\mathcal{BB}$, from prepared malware samples. Candidate BBs are ranked by triggering rate $\mathrm{TR}$ and byte length, prioritizing higher $\mathrm{TR}$ and shorter sequences. For each target benign sample, Bi-Iocane sequentially injects ranked BBs and queries the target AV until the modified sample is classified as malicious or $\mathcal{BB}$ is exhausted. A successfully flipped sample is recorded as a poisoned benign sample.

\subsubsection{Poisoned Sample Variant Generation}

A single poisoned sample may be insufficient to influence the downstream ML-based detector, so Bi-Iocane generates multiple lightweight variants using \texttt{pefile}~\cite{pefile2026} and \texttt{LIEF}~\cite{LIEF2026}. The variants are created through small random byte replacement outside BP or BB regions, attribute modification such as changing timestamps, section names, version information, or checksums, and structural modification such as adding overlays, appending sections, or shifting the optional header.
Since poisoned samples are used for training rather than execution, strict functionality preservation is not required. Nevertheless, these operations maintain valid PE headers and structural attributes, reducing interference with ML-based detectors that extract header-level features. Bi-Iocane randomly applies only one modification operation to each variant to minimize representation changes.

\subsubsection{Upload to AV-based Labeling Services}

The poisoned samples and their variants, generated using boundary knowledge from one or multiple AV engines, are uploaded to AV-based labeling services such as VirusTotal. The corresponding AV engines assign flipped labels to these samples. The uploaded samples and reports may then become accessible to authorized downstream users for dataset construction and model training.

\begin{figure}[t]\centering
	\centering
	\includegraphics[width=1\linewidth]{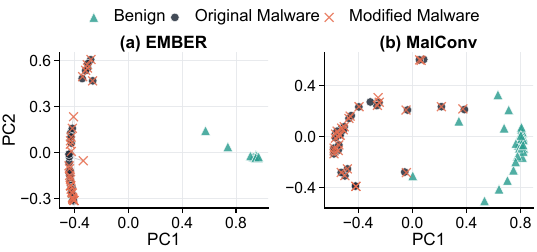}

\caption{PCA Visualization of Feature-Space Preservation after BP Rewriting.}

	\label{fig:PCA Visualization}

\end{figure}

\subsection{Poisoning Downstream ML-based Detectors}

After accessing upstream AV-based labeling services, victim dataset builders or ML detector developers collect poisoned samples and their flipped AV-assigned labels. Because these labels are treated as ground truth, the collected samples enter downstream training datasets, causing the ML-based detector to learn from corrupted sample-label pairs.

During feature extraction, the poisoned samples remain close to their original counterparts because Bi-Iocane introduces only lightweight and localized modifications. These modifications are sufficient to flip upstream AV verdicts, but the downstream detector does not possess the AV engines' pattern knowledge and may not capture the specific byte-level evidence responsible for the label reversal.

For example, in a statistical-feature detector such as EMBER~\cite{anderson2018ember}, modifying an AV-sensitive byte affects only a few histogram bins. In a raw-byte detector such as MalConv~\cite{raff2017malware}, the modification affects only a few input positions and may be diluted through embedding, convolution, and pooling. Figure~\ref{fig:PCA Visualization} visualizes 30 benign samples, 30 original malware samples, and their 30 BP-rewritten counterparts. The substantial overlap between original and BP-rewritten malware in both feature spaces provides empirical evidence that BP rewriting causes only limited representation changes.

As more poisoned variants are collected and used for training, the downstream detector is repeatedly exposed to representations close to the original targets but associated with flipped labels. This distorts the learned decision boundary around the targets and causes the trained model to misclassify the original, unmodified samples. Consequently, Bi-Iocane enables both \textit{evasion}, where malware is classified as benign, and \textit{defamation}, where benign software is classified as malicious.

\section{Evaluation}

We evaluate Bi-Iocane under a downstream-detector-agnostic black-box setting. Specifically, we assess its ability to reverse AV labels, induce targeted misclassification once the resulting samples enter downstream training data, preserve clean-set performance, and resist existing defenses. This poisoning evaluation is conditioned on attacker-submitted samples being collected by the victim's data acquisition pipeline. Our experiments address the following research questions:

\begin{itemize}

\item{\textbf{RQ1: Label Reversal Effectiveness.}} Can Bi-Iocane effectively flip AV-assigned labels with minimal byte-level perturbations?

\item{\textbf{RQ2: Targeted Poisoning Efficacy.}} Can Bi-Iocane induce targeted misclassification in downstream ML-based detectors with a low poisoning ratio?

\item{\textbf{RQ3: Model Performance Preservation.}} Does Bi-Iocane preserve the overall clean-set detection performance of poisoned ML-based detectors?

\item{\textbf{RQ4: Resistance to Poisoning Defenses.}} Can Bi-Iocane remain effective against common data sanitization and poisoning defense methods?

\end{itemize}

\subsection{Evaluation Setup}

\subsubsection{Dataset}

Malware samples are collected from BODMAS~\cite{yang2021bodmas}, SOREL-20M~\cite{harang2020sorel}, and an internal corpus derived from VirusShare~\cite{virusshare} and VirusTotal~\cite{virustotal}, while benign samples come from widely used commercial PE applications and Windows system executables. 
To reduce scanning and feature-extraction overhead, all samples are limited to 1 MB. 
In total, the dataset contains 241,070 PE files, including 120,573 malware samples and 120,497 benign samples. 
We split the dataset into training, validation, and testing sets with a ratio of 7:1:2, stratified by malware type to preserve category distribution across splits. 
The original training set is used to train clean baseline detectors, while poisoned training sets are constructed by injecting Bi-Iocane-generated samples. 
The validation set is used for model selection and early stopping, and the clean testing set is kept unchanged for evaluating attack success and clean-set performance.

\subsubsection{Upstream AV-based Labeling Services}

We evaluate Bi-Iocane under both single-AV and multi-AV labeling settings. To support large-scale automated experiments, we deploy 13 real-world AV engines that provide reliable static scanning interfaces: ClamAV~\cite{ClamAV2026}, Microsoft Defender (MS Defender)~\cite{Windows2026Defender}, Kaspersky~\cite{Kaspersky2026}, ESET~\cite{ESET2026}, Avast~\cite{Avast2026}, AVG~\cite{AVG2026}, DrWeb~\cite{DrWeb2026}, F-Secure~\cite{FSecure2026}, Vba32~\cite{Vba32}, IkarusT3~\cite{Ikarus2026}, FProtect~\cite{FProtect2012}, Avira~\cite{Avira2026}, and McAfee~\cite{McAfee2026}. Each engine individually serves as a single-AV label source.
To approximate aggregation services such as VirusTotal~\cite{virustotal}, we further combine the verdicts of all 13 engines to construct a controlled multi-AV labeling service. This setting does not reproduce VirusTotal's complete 70-plus-engine environment because of its strict query restrictions, but enables systematic and repeatable evaluation under aggregated AV labels. We additionally conduct representative validation on VirusTotal to assess the practical relevance of this setting.
All AV engines are used under their default configurations. For products without a desktop command-line interface, we use their corresponding offline command-line scanners.

\subsubsection{ML-based detectors}

We evaluate Bi-Iocane on eight representative ML-based malware detectors spanning six feature families: an EMBER-style detector based on handcrafted statistical features~\cite{anderson2018ember}; MalConv~\cite{raff2017malware} and MalConv2~\cite{raff2021classifying}, two raw-byte sequence detectors; PE-Miner~\cite{shafiq2009pe}, a PE structural-feature detector; OpcodeTransformer (OpTrans)~\cite{ren2020end}, an opcode-sequence semantic detector; MalGraph~\cite{ling2022malgraph}, a graph-based program-representation detector; and InceptionV3~\cite{szegedy2016rethinking} and IMCFN~\cite{vasan2020imcfn}, two image-based malware detectors. 
These detectors cover handcrafted static features, raw-byte modeling, PE structural analysis, instruction-level semantics, graph-based representations, and visual malware representations. 
This diverse selection allows us to evaluate whether Bi-Iocane generalizes across heterogeneous downstream feature spaces rather than exploiting a detector-specific weakness.

\begin{table*}[t]
\renewcommand{\arraystretch}{1.2}
\setlength{\tabcolsep}{4.5pt}
\caption{Boundary-Point Probing Efficiency and Malware-to-Benign Label Reversal via BP Rewriting.}
\label{tab:Malware_To_Benign}
\centering

\resizebox{\linewidth}{!}{
\begin{tabular}{ccccccccccccccc}
\hline
\textbf{Metrics}                    & \textbf{ClamAV} & \textbf{MS Defender} & \textbf{Kaspersky} & \textbf{ESET} & \textbf{Avast} & \textbf{AVG} & \textbf{DrWeb} & \textbf{FSecure} & \textbf{Vba32} & \textbf{IkarusT3} & \textbf{FProtect} & \textbf{Avira} & \textbf{McAfee} & \textbf{Average} \\ \hline
Avg. Queries Per BP (Times)         & 19.23           & 19.40             & 19.33              & 19.00         & 19.20          & 19.07        & 19.03          & 19.34            & 19.51          & 19.22             & 19.02             & 19.50          & 19.20           & 19.23            \\
Avg. Query Time Per Sample (Second) & 5.56            & 2.16              & 4.23               & 1.18          & 3.15           & 1.12         & 0.89           & 4.16             & 0.20           & 0.94              & 0.59              & 0.78           & 1.10            & 2.00             \\
Perturbation Rate (‱)               & 0.12            & 0.10              & 0.09               & 0.12          & 0.09           & 0.09         & 0.06           & 0.10             & 0.04           & 0.11              & 0.06              & 0.09           & 0.07            & 0.09             \\
LRR (\%)                            & 93.15           & 82.75             & 85.61              & 83.06         & 99.12          & 99.12        & 92.01          & 89.98            & 99.26          & 96.60             & 83.34             & 83.95          & 94.58           & 90.96            \\ \hline
\end{tabular}
}

\end{table*}

\begin{table*}[t]
\renewcommand{\arraystretch}{1.2}
\setlength{\tabcolsep}{4.5pt}
\caption{Boundary-Byte Probing Efficiency and Benign-to-Malware Label Reversal Effectiveness via BB Injection.}
\label{tab:Benign_To_Malware}
\centering

\resizebox{\linewidth}{!}{
\begin{tabular}{ccccccccccccccc}
\hline
\textbf{Metrics}                & \textbf{ClamAV} & \textbf{MS Defender} & \textbf{Kaspersky} & \textbf{ESET} & \textbf{Avast} & \textbf{AVG} & \textbf{DrWeb} & \textbf{FSecure} & \textbf{Vba32} & \textbf{IkarusT3} & \textbf{FProtect} & \textbf{Avira} & \textbf{McAfee} & \textbf{Average} \\ \hline
Avg. Queries Per BB (Times)     & 12.25           & 13.41             & 12.50              & -             & 9.43           & 9.28         & 14.00          & 10.19            & 14.00          & 9.84              & -                 & -              & 13.20           & 11.81            \\
Avg. Query Time Per BB (Second) & 69.00           & 50.72             & 858.60             & -             & 237.90         & 231.70       & 46.40          & 2573.60          & 207.10         & 299.30            & -                 & -              & 353.80          & 492.81           \\
Perturbation Rate (\%)          & 0.03            & 0.02              & 0.06               & -             & 0.01           & 0.02         & 0.64           & 0.02             & 0.18           & 0.34              & -                 & -              & 0.01            & 0.13             \\
LRR (\%)                        & 99.99           & 99.98             & 54.26              & -             & 99.99          & 99.97        & 96.16          & 79.20            & 53.54          & 96.15             & -                 & -              & 80.35           & 85.96            \\ \hline
\end{tabular}
}
\end{table*}

\subsubsection{Evaluation Metrics}

We evaluate Bi-Iocane using four metrics:

\begin{itemize}

    \item \textbf{Label Reversal Rate (LRR)} measures the fraction of attempted samples whose AV-assigned labels are successfully flipped by BP rewriting or BB injection: $\mathrm{LRR}=\frac{\#\text{label-reversed samples}}{\#\text{attempted samples}}$. 

     \item \textbf{Attack Success Rate (ASR)} measures the fraction of target samples misclassified as intended by the poisoned detector: $\mathrm{ASR}=\frac{\#\text{successful target misclassifications}}{\#\text{target samples}}$. 
    
     \item \textbf{Poisoning Ratio (PR)} denotes the proportion of poisoned samples in the training set: $\mathrm{PR}=\frac{|\mathcal{D}_{\mathrm{poison}}|}{|\mathcal{D}_{\mathrm{clean}}|+|\mathcal{D}_{\mathrm{poison}}|}$. 
    
        \item \textbf{Clean-set Performance} is measured by precision, recall, and F1 on the clean test set: $P=\frac{TP}{TP+FP}$, $R=\frac{TP}{TP+FN}$, $\mathrm{F1}=\frac{2PR}{P+R}$.

\end{itemize}

\subsubsection{Experimental Setup}

All experiments are conducted on a Windows 11 system equipped with an NVIDIA GeForce RTX 4080 SUPER GPU and an Intel Core i7-14700K CPU. 
For each target detector, we train a clean baseline model on the original training set using the recommended configuration from its original implementation. 
Each model is trained for at most 100 epochs, with early stopping based on validation performance. 
The checkpoint with the best validation performance is selected as the final clean baseline. 
The clean-set F1 scores of these baseline detectors are reported in Table~\ref{tab:F1_Performance}.
For poisoning evaluation, Bi-Iocane-generated samples are injected into the training set, and the corresponding detector is retrained from scratch. 
To ensure fair comparison, each poisoned model uses the same architecture, hyperparameters, data split, and maximum number of training epochs as its clean baseline. 
The clean test set is kept unchanged and used to evaluate attack success and clean-set performance, including precision, recall, and F1-score.

\subsection{RQ1: Label Reversal Effectiveness}

\subsubsection{Malware-to-Benign Label Reversal via BP Rewriting}

We evaluate whether Bi-Iocane can reverse AV-assigned malware labels through BP rewriting. For each AV engine, Bi-Iocane probes the BPs of malware samples in the test set and masks the identified bytes with \texttt{0x90}. We limit the number of probed BPs to 20 per sample for tractability. Samples reaching this limit are conservatively treated as failures because their AV decisions may depend on complex or dispersed patterns. Table~\ref{tab:Malware_To_Benign} reports the results across 13 AV engines.

Bi-Iocane achieves an average LRR of 90.96\%, demonstrating that malware labels can be reversed by modifying only a few boundary bytes. The LRRs reach 99.26\% for Vba32 and 99.12\% for both Avast and AVG, while even the lowest LRR, observed on Microsoft Defender, remains 82.75\%. The average perturbation rate is only 0.09\textpertenthousand. BP probing requires 19.23 queries per BP and 2.00 seconds per sample on average.

\subsubsection{Benign-to-Malware Label Reversal via BB Injection}

We next evaluate benign-to-malware label reversal through BB injection. Bi-Iocane extracts transferable BBs around the probed BPs and retains at most 100 effective BBs per AV engine whose triggering rates exceed the predefined threshold. These BBs are injected into benign test samples following Section~\ref{Section:Poisoned Benign Sample Generation}.

As shown in Table~\ref{tab:Benign_To_Malware}, Bi-Iocane identifies effective BBs for 10 of the 13 AV engines and achieves an average LRR of 85.96\%. ClamAV, Microsoft Defender, Avast, and AVG exhibit LRRs close to 100\%, while DrWeb and IkarusT3 exceed 96\%. The lower LRRs of Kaspersky and Vba32 suggest a stronger dependence on contextual or combined patterns.
No effective BB is identified for ESET, FProtect, or Avira under the current probing budget, possibly because a single BB is insufficient to trigger their malicious verdicts. The averages are therefore computed over the 10 AV engines with effective BBs. BB probing requires 492.81 seconds per BB on average because not every BP context is transferable. Nevertheless, its average perturbation rate is only 0.13\%, enabling effective label reversal for most AV engines with limited modifications.

\subsubsection{Real-World Validation on a Multi-AV Aggregation Platform}
\label{section:Real-World Validation on a Multi-AV Aggregation Platform}
To assess Bi-Iocane on a real-world multi-AV aggregation platform, we randomly select 30 BP-rewritten malware samples and 30 BB-injected benign samples from the test set and submit their original and modified versions to VirusTotal.

For benign samples, BB injection increases the average number of AV detections from 0 to 25.90. Since AV-count-based dataset construction commonly treats samples detected by approximately 10 or more engines as malicious, these modified samples can receive malicious labels and enter downstream training datasets, demonstrating the practical feasibility of defamation-oriented poisoning.
For malware samples, BP rewriting reduces the average number of detections from 52.40 to 29.43, a decrease of 22.97 engines. This result does not constitute complete malware-to-benign reversal because benign labels generally require no or only a few AV detections. However, it shows that BP rewriting can remove detection evidence from a substantial number of engines.

A complete evaluation is limited by VirusTotal's query restrictions. BP rewriting introduces only 0.09\textpertenthousand{} perturbation per AV engine on average. As a rough feasibility estimate, rewriting BPs for approximately 70 engines would introduce about 6.3\textpertenthousand{}, or 0.063\%, perturbation. Although this estimate does not prove complete evasion on VirusTotal, it suggests that extending BP rewriting to more engines may be feasible without substantially changing downstream ML representations.

\begin{tcolorbox}[
colback=gray!20,
colframe=black!50,
boxrule=0.4pt,
arc=1mm,
left=1.5mm,
right=1.5mm,
top=1mm,
bottom=1mm,
boxsep=0mm,
before skip=2pt,
after skip=2pt,
fontupper=\footnotesize
]
\textbf{Answer to RQ1}: 
Bi-Iocane achieves effective bidirectional label reversal with minimal byte-level perturbations. On VirusTotal, BB injection crosses common malicious-label thresholds, while BP rewriting substantially reduces AV detections, indicating potential for broader multi-engine evasion.

\end{tcolorbox}

\begin{figure*}[t]\centering
	\centering
	\includegraphics[width=1\linewidth]{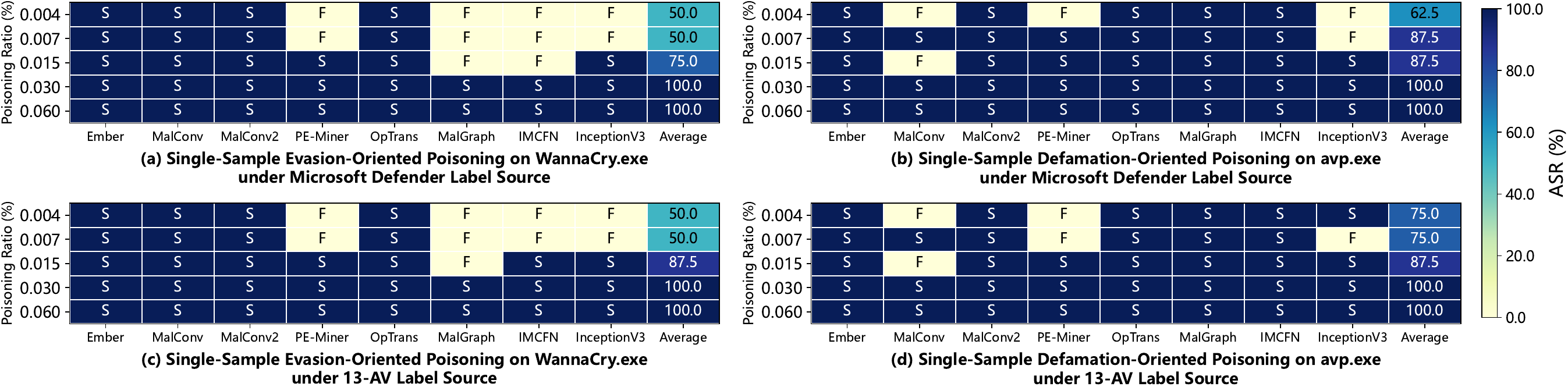}
\caption{Single-sample poisoning results under the Microsoft Defender and 13-AV label sources. 
Each cell indicates whether Bi-Iocane successfully induces the intended misclassification at a given poisoning ratio.}
    
	\label{fig:Single_Sample_Poisoning}
    
\end{figure*}

\begin{table*}[t]

\renewcommand{\arraystretch}{1.2}
\setlength{\tabcolsep}{4.5pt}
\caption{Batch-Target Boundary Label Poisoning Results for Evasion and Defamation on 30 Malware and 30 Benign Samples Under Microsoft Defender and 13-AV Labels.}
\label{tab:Batch_Target_Poisoning}
\centering

\resizebox{1\linewidth}{!}{
\begin{tabular}{cccccccccccccc}
\hline
\textbf{Label Source}               & \textbf{Per-target PR} & \textbf{Overall PR} & \textbf{Ember} & \textbf{Malconv} & \textbf{Malconv2} & \textbf{PE-Miner} & \textbf{OpTrans} & \textbf{MalGraph} & \textbf{IMCFN} & \textbf{InceptionV3} & \textbf{Evasion Avg.} & \textbf{Defamation Avg.} & \textbf{Overall Avg.} \\ \hline
\multirow{5}{*}{Microsoft Defender} & 0.004                  & 0.220               & 100.00         & 68.33            & 70.00             & 46.67             & 75.00            & 41.67             & 91.67          & 23.33                & 54.58                 & 74.58                    & 64.58                 \\
                                    & 0.007                  & 0.404               & 96.67          & 88.33            & 75.00             & 90.00             & 91.67            & 63.33             & 93.33          & 40.00                & 78.75                 & 80.83                    & 79.79                 \\
                                    & 0.015                  & 0.955               & 98.33          & 88.33            & 96.67             & 100.00            & 88.33            & 71.67             & 100.00         & 81.67                & 92.50                 & 88.75                    & 90.63                 \\
                                    & 0.030                  & 1.874               & 100.00         & 86.67            & 91.67             & 100.00            & 96.67            & 91.67             & 100.00         & 91.67                & 93.75                 & 95.83                    & 94.79                 \\
                                    & 0.060                  & 3.711               & 100.00         & 93.33            & 100.00            & 100.00            & 98.33            & 98.33             & 100.00         & 93.33                & 98.36                 & 97.47                    & 97.92                 \\ \hline
\multirow{5}{*}{13-AV}              & 0.004                  & 0.220               & 76.67          & 50.00            & 48.33             & 38.33             & 80.00            & 48.33             & 85.00          & 23.33                & 54.17                 & 58.33                    & 56.25                 \\
                                    & 0.007                  & 0.404               & 81.67          & 75.00            & 61.67             & 68.33             & 86.67            & 50.00             & 96.67          & 43.33                & 69.58                 & 71.25                    & 70.42                 \\
                                    & 0.015                  & 0.955               & 88.33          & 73.33            & 96.67             & 85.00             & 95.00            & 66.67             & 100.00         & 51.67                & 83.73                 & 80.42                    & 82.08                 \\
                                    & 0.030                  & 1.874               & 90.00          & 81.67            & 70.00             & 85.00             & 83.33            & 76.67             & 98.33          & 81.67                & 85.41                 & 81.25                    & 83.33                 \\
                                    & 0.060                  & 3.711               & 93.33          & 88.33            & 86.67             & 90.00             & 100.00           & 90.00             & 100.00         & 88.33                & 96.66                 & 87.50                    & 92.08                 \\ \hline
\end{tabular}
}

\vspace{1.0mm}
\begin{minipage}{0.98\linewidth}
\scriptsize

\emph{Notes.} All detector-specific results and averages denote ASR (\%). 
Evasion Avg. and Defamation Avg. are averaged over 30 malware targets and 30 benign targets, respectively, across eight detectors, while Overall Avg. is computed over all 60 targets. 
Per-target PR denotes the poisoning ratio for each target, whereas Overall PR denotes the total poisoning ratio after aggregating all 60 targets.

\end{minipage}

\end{table*}

\subsection{RQ2: Targeted Poisoning Efficacy}

RQ2 evaluates whether the label-flipped samples validated in RQ1 can induce targeted misclassification in downstream ML-based malware detectors under different poisoning ratios. We consider both single-sample and batch-target settings. For the single-sample setting, we select one representative malware sample, \texttt{WannaCry.exe}\footnote{\fontsize{6.5pt}{7.5pt}\selectfont SHA256: ed01ebfbc9eb5bbea545af4d01bf5f1071661840480439c6e5babe8e080e41aa}, and one representative benign program, Kaspersky's main scanning process \texttt{avp.exe}\footnote{\fontsize{6.5pt}{7.5pt}\selectfont SHA256: f086ad27cd5644b6d214cb4426efdef999d02419eda78ea6e7c90293ece596dc}. For the batch-target setting, we use 60 target samples, including 30 malware and 30 benign samples from the real-world multi-AV validation in Section~\ref{section:Real-World Validation on a Multi-AV Aggregation Platform}. We use a limited number of targets because each target requires multiple poisoned variants, and excessive poisoning would unrealistically distort the training distribution and affect clean-set evaluation. This setting better reflects practical targeted poisoning, where the adversary aims to manipulate a small set of selected samples. Before poisoning, all targets are correctly labeled by the corresponding AV label sources and correctly classified by the clean ML-based detectors.

For each target, Bi-Iocane generates one label-flipped sample and lightweight variants. We inject 5, 10, 25, 50, and 100 poisoned instances per target, yielding per-target PRs of 0.004\%, 0.007\%, 0.015\%, 0.030\%, and 0.060\%, and overall PRs of 0.220\%, 0.404\%, 0.955\%, 1.874\%, and 3.711\% for the 60 batch targets. Evasion targets original malware, whereas defamation targets original benign software.

\subsubsection{Single-AV Label Source}

We first use Microsoft Defender as the single upstream label source and inject the Defender-derived poisoned samples and their variants into the training set under the five poisoning ratios.

As shown in Figure~\ref{fig:Single_Sample_Poisoning}, the average ASRs of \texttt{WannaCry.exe} and \texttt{avp.exe} increase from 50.0\% and 62.5\% at the lowest PR to 100.0\% at the highest PR, respectively. Table~\ref{tab:Batch_Target_Poisoning} shows that the overall average ASR across the 60 batch targets increases from 64.58\% to 97.92\%. Over the same range, the average evasion and defamation ASRs increase from 54.58\% and 74.58\% to 98.36\% and 97.47\%, respectively. At a per-target PR of only 0.060\%, six of the eight detectors achieve ASRs of at least 98\%, including 100.0\% on EMBER, MalConv2, PE-Miner, and IMCFN. These results demonstrate effective targeted poisoning on both representative and randomly selected targets under a single-AV label source.

\subsubsection{Multi-AV Label Source}

We next use the 13 evaluated AV engines to construct a controlled multi-AV aggregation label source, while keeping the targets, poisoning ratios, and downstream detectors unchanged. For each target, AV-specific boundary manipulations are combined within the same executable. Evasion rewrites the BPs identified by the evaluated AV engines, while defamation injects the available effective AV-specific BBs. We adopt a conservative aggregation rule: a sample is labeled benign only if all 13 engines report benign; otherwise, any malicious verdict yields a malicious label.

Under the 13-AV label source, the average ASRs of \texttt{WannaCry.exe} and \texttt{avp.exe} increase from 50.0\% and 75.0\% at the lowest PR to 100.0\% at the highest PR, respectively. For the batch targets, the overall average ASR increases from 56.25\% to 92.08\%. At a per-target PR of 0.060\%, evasion and defamation achieve average ASRs of 96.66\% and 87.50\%, respectively, and five detectors achieve ASRs of at least 90\%.

Although aggregating multiple AV verdicts makes label manipulation more restrictive, Bi-Iocane remains effective across all six downstream feature families. The lower defamation ASR under the 13-AV setting is likely caused by the larger combined BB perturbation, which may make the poisoned benign samples more distinguishable from their original targets in downstream feature spaces and weaken the intended label--representation conflict.

\subsubsection{Practical Poisoning Implication Analysis}

Bi-Iocane becomes more effective as additional poisoned variants are introduced, supporting the underlying mechanism that repeated label--representation conflicts gradually shift the decision boundary around target samples. Its effectiveness on both single and batch targets further indicates that the attack is not limited to isolated cases.

Combined with VirusTotal validation in Section~\ref{section:Real-World Validation on a Multi-AV Aggregation Platform}, the results demonstrate a practical defamation risk to ML pipelines relying on multi-AV aggregated labels. Evasion also remains effective under both single-AV and controlled multi-AV settings. Although the complete VirusTotal environment cannot be reproduced, these findings reveal a potential poisoning risk for ML pipelines using upstream aggregation providers.

\begin{tcolorbox}[
colback=gray!20,
colframe=black!50,
boxrule=0.4pt,
arc=1mm,
left=1.5mm,
right=1.5mm,
top=1mm,
bottom=1mm,
boxsep=0mm,
before skip=2pt,
after skip=2pt,
fontupper=\footnotesize
]
\textbf{Answer to RQ2}:
Bi-Iocane successfully induces targeted evasion and defamation across heterogeneous ML-based detectors under both single-AV and multi-AV labeling settings. Its effectiveness across different feature families and poisoning ratios demonstrates that the attack is not limited to a specific downstream detector.

\end{tcolorbox}

\begin{table*}[t]
\renewcommand{\arraystretch}{1}
\setlength{\tabcolsep}{4.5pt}
\caption{Clean-Set F1 Performance of Poisoned Models under Different Poisoning Settings.}
\label{tab:F1_Performance}
\centering

\resizebox{\linewidth}{!}{
\begin{tabular}{ccccccccccc}
\hline
\textbf{Label Source}                & \textbf{Target Setting}         & \textbf{Ember} & \textbf{MalConv} & \textbf{MalConv2} & \textbf{PE-Miner} & \textbf{OpTrans} & \textbf{MalGraph} & \textbf{IMCFN} & \textbf{InceptionV3} & \textbf{Average} \\ \hline
                                     & Single-Sample Evasion           & 99.17$\pm$0.03 & 98.37$\pm$0.06   & 98.62$\pm$0.10    & 98.86$\pm$0.01    & 93.00$\pm$0.23   & 96.08$\pm$0.07    & 97.02$\pm$0.19 & 97.12$\pm$0.20       & 97.28            \\
                                     & Single-Sample Defamation        & 99.16$\pm$0.03 & 98.66$\pm$0.05   & 98.77$\pm$0.08    & 98.87$\pm$0.01    & 92.94$\pm$0.26   & 96.24$\pm$0.06    & 96.99$\pm$0.17 & 97.13$\pm$0.19       & 97.34            \\
\multirow{-3}{*}{Microsoft Defender} & Batch-Target Evasion+Defamation & 99.12$\pm$0.04 & 98.31$\pm$0.06   & 98.59$\pm$0.06    & 98.85$\pm$0.02    & 94.10$\pm$1.62   & 96.03$\pm$0.16    & 96.92$\pm$0.13 & 96.85$\pm$0.22       & 97.35            \\ \hline
                                     & Single-Sample Evasion           & 99.16$\pm$0.03 & 98.44$\pm$0.21   & 98.79$\pm$0.10    & 98.86$\pm$0.01    & 92.96$\pm$0.38   & 96.15$\pm$0.13    & 96.95$\pm$0.16 & 97.18$\pm$0.21       & 97.31            \\
                                     & Single-Sample Defamation        & 99.15$\pm$0.04 & 98.69$\pm$0.06   & 98.73$\pm$0.05    & 98.86$\pm$0.01    & 93.10$\pm$0.46   & 96.03$\pm$0.15    & 97.07$\pm$0.04 & 97.15$\pm$0.08       & 97.35            \\
\multirow{-3}{*}{13-AV}              & Batch-Target Evasion+Defamation & 99.14$\pm$0.03 & 98.58$\pm$0.07   & 98.69$\pm$0.10    & 98.85$\pm$0.02    & 93.96$\pm$0.39   & 95.97$\pm$0.13    & 97.07$\pm$0.15 & 96.73$\pm$0.28       & 97.37            \\ \hline
\rowcolor[HTML]{EFEFEF} 
\multicolumn{2}{c}{\cellcolor[HTML]{EFEFEF}Clean Baseline}             & 99.17          & 98.71            & 98.64             & 98.87             & 93.48            & 96.39             & 97.00          & 97.15                & 97.43            \\ \hline
\end{tabular}
}

\vspace{1.0mm}
\begin{minipage}{0.98\linewidth}
\scriptsize
\emph{Notes.} Values denote clean-set F1 scores (\%). The Clean Baseline row reports models trained on the original clean training set. For each poisoned setting, mean$\pm$std is computed over the five poisoning-ratio settings. 
\end{minipage}

\end{table*}

    
    

\begin{table*}[t]
\renewcommand{\arraystretch}{1}
\setlength{\tabcolsep}{4.5pt}
\caption{Attack Success Rate of Bi-Iocane after Applying Different Defense Methods.}
\label{tab:defense_resistance}
\centering

\resizebox{0.85\linewidth}{!}{
\begin{tabular}{ccccccccccc}
\hline
\textbf{Label Source}                & \textbf{Defense Method}                       & \textbf{Ember}                 & \textbf{MalConv}              & \textbf{MalConv2}              & \textbf{PE-Miner}              & \textbf{OpTrans}               & \textbf{MalGraph}             & \textbf{IMCFN}                 & \textbf{InceptionV3}          & \textbf{Average}              \\ \hline

                                     & Spectral Signature~\cite{tran2018spectral}                            & 98.33                          & 93.33                         & 100.00                         & 96.67                          & 86.67                          & 93.33                         & 91.67                          & 93.33                         & 93.13                         \\
                                     & Sphere/Slab Filtering~\cite{steinhardt2017certified}                         & 100.00                         & 83.33                         & 90.00                          & 91.67                          & 86.67                          & 100.00                        & 93.33                          & 93.33                         & 90.42                         \\
                                     & Beatrix~\cite{ma2022beatrix}                                       & 88.33                          & 88.33                         & 91.67                          & 91.67                          & 88.33                          & 85.00                         & 93.33                          & 95.00                         & 88.75                         \\
                                     & Loss-Based Filtering~\cite{li2021anti}                          & 100.00                         & 81.67                         & 90.00                          & 100.00                         & 85.00                          & 81.67                         & 88.33                          & 93.33                         & 88.54                         \\
                                     & SCAn~\cite{tang2021demon}                                          & 100.00                         & 91.67                         & 96.67                          & 100.00                         & 81.67                          & 85.00                         & 98.33                          & 98.33                         & 92.50                         \\
                                     & DPA~\cite{levine2020deep}                                           & 99.67                          & 84.67                         & 97.67                          & 100.00                         & 79.67                          & 77.33                         & 90.00                          & 94.00                         & 88.88                         \\
\multirow{-7}{*}{Microsoft Defender} & \cellcolor[HTML]{EFEFEF}Baseline (No Defense) & \cellcolor[HTML]{EFEFEF}100.00 & \cellcolor[HTML]{EFEFEF}93.33 & \cellcolor[HTML]{EFEFEF}100.00 & \cellcolor[HTML]{EFEFEF}100.00 & \cellcolor[HTML]{EFEFEF}98.33  & \cellcolor[HTML]{EFEFEF}98.33 & \cellcolor[HTML]{EFEFEF}100.00 & \cellcolor[HTML]{EFEFEF}93.33 & \cellcolor[HTML]{EFEFEF}97.92

\\ \hline

  & Spectral Signature~\cite{tran2018spectral}                            & 83.33                          & 73.33                         & 91.67                          & 88.33                          & 88.33                          & 91.67                         & 98.33                          & 81.67                         & 86.67                         \\
                                     & Sphere/Slab Filtering~\cite{steinhardt2017certified}                         & 83.33                          & 70.00                         & 81.67                          & 76.67                          & 88.33                          & 90.00                         & 100.00                         & 90.00                         & 86.19                         \\
                                     & Beatrix~\cite{ma2022beatrix}                                       & 83.33                          & 70.00                         & 83.33                          & 80.00                          & 88.33                          & 91.67                         & 96.67                          & 83.33                         & 83.10                         \\
                                     & Loss-Based Filtering~\cite{li2021anti}                          & 90.00                          & 71.67                         & 88.33                          & 88.33                          & 93.33                          & 85.00                         & 55.00                          & 91.67                         & 82.14                         \\
                                     & SCAn~\cite{tang2021demon}                                          & 88.33                          & 76.67                         & 83.33                          & 88.33                          & 91.67                          & 66.67                         & 98.33                          & 85.00                         & 84.29                         \\
                                     & DPA~\cite{levine2020deep}                                           & 84.00                          & 79.33                         & 95.00                          & 88.00                          & 86.00                          & 73.67                         & 99.67                          & 87.00                         & 86.38                         \\
\multirow{-7}{*}{13-AV}             & \cellcolor[HTML]{EFEFEF}Baseline (No Defense) & \cellcolor[HTML]{EFEFEF}93.33  & \cellcolor[HTML]{EFEFEF}88.33 & \cellcolor[HTML]{EFEFEF}86.67  & \cellcolor[HTML]{EFEFEF}90.00  & \cellcolor[HTML]{EFEFEF}100.00 & \cellcolor[HTML]{EFEFEF}90.00 & \cellcolor[HTML]{EFEFEF}100.00 & \cellcolor[HTML]{EFEFEF}88.33 & \cellcolor[HTML]{EFEFEF}92.08

\\ \hline
\end{tabular}
}

\vspace{1.0mm}
\begin{minipage}{0.98\linewidth}
\scriptsize

\emph{Notes.} Values denote ASR (\%) after each defense. For DPA, ASR is averaged over five subsets, where each subset-level ASR is computed as the proportion of original target samples that remain successfully attacked after DPA ensemble. Other defenses report ASR after removing the top 5\% suspicious samples.

\end{minipage}

\end{table*}

\subsection{RQ3: Model Performance Preservation}

RQ3 evaluates whether the poisoned models obtained in RQ2 preserve the clean-set detection performance of downstream ML-based malware detectors. 
Although RQ2 shows that Bi-Iocane can induce targeted evasion and defamation, a practical poisoning attack should not substantially degrade the detector's overall performance on clean samples. 
Otherwise, an abnormal performance drop may reveal the poisoning effect during model validation or deployment.

To evaluate this, we directly test the poisoned models trained in RQ2 on the unchanged clean test set. 
These models cover two AV label sources, Microsoft Defender and 13-AV, and three target settings under each label source: single-sample evasion, single-sample defamation, and batch-target evasion plus defamation. 
For each poisoned setting, we summarize the results over the five per-target PRs used in RQ2. 
Table~\ref{tab:F1_Performance} reports the per-detector clean-set F1 scores, where each poisoned setting is reported as mean$\pm$std over the five poisoning ratios. 

The results show that Bi-Iocane has only a limited impact on clean-set performance. 
The clean baseline achieves an average F1 score of 97.43\%. 
Under the Microsoft Defender label source, the average F1 scores of the poisoned models are 97.28\%, 97.34\%, and 97.35\% for single-sample evasion, single-sample defamation, and batch-target evasion plus defamation, respectively. 
These values correspond to decreases of only 0.15, 0.09, and 0.08 percentage points from the clean baseline. 
Under the 13-AV label source, the average F1 scores are 97.31\%, 97.35\%, and 97.37\%, with decreases of only 0.12, 0.08, and 0.06 percentage points, respectively.

At the per-detector level, Table~\ref{tab:F1_Performance} shows that clean-set F1 scores remain stable across different detector architectures and feature families. 
For example, EMBER, PE-Miner, MalConv2, IMCFN, and InceptionV3 all maintain F1 scores close to their clean baselines across poisoned settings. 
Although some detectors show small fluctuations under certain settings, the overall variation remains minor and does not indicate systematic degradation on clean samples. 
This suggests that Bi-Iocane mainly affects the decision behavior around selected target samples, while preserving the general detection capability of poisoned models on clean data.

\begin{tcolorbox}[
colback=gray!20,
colframe=black!50,
boxrule=0.4pt,
arc=1mm,
left=1.5mm,
right=1.5mm,
top=1mm,
bottom=1mm,
boxsep=0mm,
before skip=2pt,
after skip=2pt,
fontupper=\footnotesize
]
\textbf{Answer to RQ3}: 
Bi-Iocane preserves the overall clean-set performance of downstream ML-based malware detectors. 
Across the poisoned models obtained in RQ2, the average F1 scores remain close to the clean baseline, indicating that Bi-Iocane can induce targeted evasion and defamation without substantially degrading normal clean-sample detection performance.
\end{tcolorbox}

\subsection{RQ4: Resistance to Poisoning Defenses}

RQ4 evaluates whether existing poisoning and backdoor defenses can mitigate Bi-Iocane. We reuse the batch-target poisoning settings from RQ2 and select the highest per-target PR setting under each label source, namely 13-AV and Microsoft Defender. These settings produce stable poisoning effects and are suitable for testing whether defenses can weaken the attack.

We use No Defense as the baseline, where the poisoned training set is directly used to retrain the detector, corresponding to the RQ2 results. We then evaluate six representative defenses: Spectral Signature~\cite{tran2018spectral}, Sphere/Slab Filtering~\cite{steinhardt2017certified}, Beatrix~\cite{ma2022beatrix}, Loss-Based Filtering~\cite{li2021anti}, SCAn~\cite{tang2021demon}, and DPA~\cite{levine2020deep}. For filtering-based defenses, we remove the top 5\% suspicious samples from each class. For DPA, we use an ensemble with $k=5$ data partitions. Each defense is applied in the feature space of the corresponding detector, and ASR is measured on the original batch-target samples after defense and retraining.

Table~\ref{tab:defense_resistance} reports the defended ASR of Bi-Iocane. Overall, existing defenses provide only limited mitigation. Under the 13-AV label source, the No Defense baseline achieves an average ASR of 92.08\%, while defended models still retain average ASRs between 82.14\% and 86.67\%. Under the Microsoft Defender label source, the baseline ASR is 97.92\%, and the defended ASRs remain between 88.54\% and 93.13\%. Thus, none of the evaluated defenses consistently eliminates the targeted poisoning effect.

At the detector level, some defenses reduce ASR in specific cases. For example, under the 13-AV label source, SCAn reduces the ASR of MalGraph to 66.67\%, and Loss-Based Filtering reduces the ASR of IMCFN to 55.00\%. Under Microsoft Defender, DPA reduces the ASR of MalGraph to 77.33\%. However, most detector-defense combinations still maintain high ASR close to the No Defense baseline, indicating that the overall mitigation effect remains limited.

This limited effectiveness stems from the assumptions of existing defenses. Most defenses expect poisoned samples to exhibit separable abnormal patterns, such as outlier representations, abnormal loss behavior, or consistent trigger-induced signatures. In contrast, Bi-Iocane manipulates labels through upstream AV boundary decisions. Its poisoned samples are valid PE files and remain close to normal samples in downstream feature spaces, making them difficult for standard filtering and ensemble-based defenses to distinguish from normal training data.

\begin{tcolorbox}[
colback=gray!20,
colframe=black!50,
boxrule=0.4pt,
arc=1mm,
left=1.5mm,
right=1.5mm,
top=1mm,
bottom=1mm,
boxsep=0mm,
before skip=2pt,
after skip=2pt,
fontupper=\footnotesize
]
\textbf{Answer to RQ4}: 
Existing poisoning and backdoor defenses provide only limited protection against Bi-Iocane. 
Although a few defenses reduce ASR in specific cases, Bi-Iocane maintains high ASR for most detector-defense combinations, showing its resistance to existing defenses.
\end{tcolorbox}

\section{Related Work}
\label{Section:Related_Work}

\textbf{ML-based Malware Detectors.}
Static ML-based malware detectors have been widely studied with diverse input representations. Existing approaches can be broadly grouped into handcrafted statistical-feature detectors, which use attributes such as byte distributions, entropy, strings, imports, and metadata~\cite{anderson2018ember,mahmud2025novel,tian2025density}; raw-byte detectors, which learn directly from binary byte streams~\cite{raff2017malware,raff2021classifying}; PE structural detectors, which rely on headers, section layouts, data directories, and import/export structures~\cite{shafiq2009pe,hu2022generating}; instruction-level detectors, which analyze opcode sequences or disassembly semantics~\cite{ren2020end,mclaughlin2017deep}; graph-based detectors, which model control-flow graphs, call graphs, or other program relations~\cite{ling2022malgraph,peng2024malgne,yan2019classifying}; and image-based detectors, which convert binaries into visual representations~\cite{szegedy2016rethinking,vasan2020imcfn,ni2018malware}. These categories cover the major static feature spaces considered in this work.

\textbf{Poisoning Attacks Against ML-based Malware Detectors.}
Data poisoning attacks against ML-based malware detectors are commonly divided into \textit{dirty-label} and \textit{clean-label} attacks. Dirty-label attacks directly flip sample labels to implant backdoors or induce targeted misclassification~\cite{chen2018automated,sasaki2019embedding,narisada2020stronger,li2021backdoor,noppel2023disguising}. Clean-label attacks preserve original labels but perturb feature-relevant content, such as API calls, section data, or file headers, to poison detectors across platforms and architectures~\cite{shapira2020being,severi2021explanation,zheng2022clean,yang2023jigsaw,d2023lookin,zhang2023universal,tian2023sparsity,zhan2025practical}.

\textbf{Positioning of Bi-Iocane.}
Most existing attacks target downstream ML detectors and are designed around their feature representations or input spaces, requiring the adversary to know, infer, or approximate the target feature space. This assumption is difficult in black-box settings. AndroVenom~\cite{lan2026trust} shifts poisoning to the upstream AV-based labeling stage, but still depends on a known downstream blind spot by injecting payloads into APK resource locations ignored by common Android feature extractors. It also focuses only on unidirectional \textit{defamation}. In contrast, Bi-Iocane exploits AV-to-ML boundary misalignment to provide a downstream-detector-agnostic black-box poisoning framework that supports both \textit{evasion} and \textit{defamation} without requiring knowledge of the target architecture, training data, or feature representation.

\section{Discussion}

\subsection{Mitigations}

Bi-Iocane exploits the trust placed in upstream AV labeling services, and several data-curation practices can help reduce its impact. First, multi-AV aggregation should not be treated as a sufficient defense, since aggregated labels can also be manipulated. Dataset builders should record sample provenance, distinguish public submissions from trusted internal corpora or curated feeds, and use multi-AV verdicts only as one labeling signal. Second, because Bi-Iocane amplifies poisoning through variants derived from the same base sample, training pipelines should identify highly similar PE files using cryptographic hashes, fuzzy hashes, PE metadata, structural features, or learned representations, and limit their number or training weight. Third, samples with low AV agreement, labels near aggregation thresholds, or abrupt label changes should receive additional semantic verification, such as sandbox execution, dynamic analysis, manual inspection, or trusted internal labels, before entering downstream training data.

\subsection{Limitations and Future Work}

Despite the findings presented in this paper, several limitations remain, and motivate future work.
First, our evaluation focuses on the Windows AV-to-ML labeling pipeline. We consider Windows AV engines and PE malware detectors, but do not examine whether Bi-Iocane transfers directly to Android, Linux, or other ecosystems. Future work can extend the study to APKs, ELF binaries, and platform-specific labeling pipelines.
Second, our controlled multi-AV setting uses 13 AV engines to approximate an aggregation service such as VirusTotal. Although our VirusTotal experiments confirm the practical feasibility of defamation poisoning, BP rewriting only reduces the number of malware detections and does not consistently achieve complete evasion. Larger-scale evaluations on real multi-engine services are therefore needed to further validate evasion-oriented poisoning and study the effects of engine diversity, aggregation rules, and label thresholds.
Third, our threat model assumes that attacker-submitted samples and their AV labels can enter downstream training sets. Real dataset construction pipelines may apply provenance checks, duplicate filtering, temporal rescanning, or manual review. Future work should evaluate Bi-Iocane under such data-governance mechanisms and develop defenses tailored to AV-labeled training pipelines.

\section{Conclusions}

In this paper, we presented Bi-Iocane, a downstream-detector-agnostic black-box poisoning framework that exploits decision-boundary misalignment between AV engines and ML-based malware detectors. By rewriting malware boundary points and injecting boundary bytes into benign software, Bi-Iocane flips AV-assigned labels while largely preserving downstream ML representations.
Experiments show that Bi-Iocane enables both evasion and defamation under single-AV and multi-AV label sources, with limited impact on clean-set performance and continued effectiveness against existing defenses. VirusTotal validation confirms the practical feasibility of defamation poisoning, while the substantial reduction in malware detections reveals a potential evasion-oriented risk requiring further study.
These findings expose a critical weakness in AV-to-ML labeling pipelines and motivate more robust label generation and AV-boundary-aware defenses.

\bibliographystyle{IEEEtran}
\bibliography{BiIocane}

\end{document}